\documentclass[twocolumn,superscriptaddress,prx,aps]{revtex4-2}
\usepackage{amsmath}
\usepackage{amssymb}
\DeclareUnicodeCharacter{2212}{\textendash}
\usepackage{graphicx}
\usepackage{dcolumn}
\usepackage{bm}
\usepackage[normalem]{ulem}
\usepackage{xcolor} 
\usepackage[version=4]{mhchem} 
\usepackage{graphicx}
\usepackage{amsmath}
\usepackage{nicematrix}
\usepackage{textcomp}
\usepackage{gensymb} 
\usepackage[bb=boondox]{mathalfa}
\usepackage{subfiles}
\usepackage{tabularx} 
\usepackage{overpic}
\usepackage[linktocpage=true,
  colorlinks=true, 
  pdfborder={0 0 0}, 
  linkcolor=blue,
  citecolor=red,
  filecolor=yellow,
  urlcolor=blue,
  bookmarks,
  pdfauthor={},
]{hyperref}

\usepackage{chemformula}

\usepackage{subfiles}

\newcommand\pc[1]{\textcolor{blue}{#1}}

\begin{document}

\title{Inverse Lieb Materials: Altermagnetism and More}

\author{Po-Hao Chang}
\affiliation{George Mason University, Department of Physics and Astronomy} \affiliation{Quantum Science and Engineering Center,  Fairfax, USA}

\author{Igor I. Mazin}
\email{imazin2@gmu.edu}
\affiliation{George Mason University, Department of Physics and Astronomy} \affiliation{Quantum Science and Engineering Center,  Fairfax, USA}

\author{Kirill D. Belashchenko}
\affiliation{Department of Physics and Astronomy and Nebraska Center for Materials and Nanoscience,
University of Nebraska-Lincoln, Lincoln, Nebraska 68588, USA}

\begin{abstract}
The Lieb lattice, originally proposed for cuprate
superconductors, has gained new attention in the emerging field of altermagnetism as a minimal analytical model for the latter. While 
initially the so-called inverse Lieb lattice (ILL) was deemed to be only a convenient theoretical model, recently several real materials with this crystallographic motif have been found. The unique
geometry of ILL can accommodate complex magnetic orderings
arising from competing exchange interactions and geometric frustration,
offering great tunability for magnetic properties. In this work, we
provide comprehensive insights into magnetic phases in ILL materials 
and establish guidelines for the efficient identification of altermagnetic materials within this family. 
We begin by constructing phase diagrams
using a simple Heisenberg model to elucidate the fundamental mechanisms
underlying altermagnetism and other complex magnetic phases observed
experimentally. To bridge theory with experiment, we systematically
investigate a series of existing ILL compounds using density
functional theory (DFT) calculations to determine their magnetic ground
states. Our computational results are in good agreement with experimental
observations. 
Importantly, we identify a trend linking magnetic ordering
to the $d$-shell filling of transition metal ions, with $d^{2-3}$
and $d^{5}$ configurations showing propensity for altermagnetic behavior.
Additionally, we identify a promising metallic compound Sr$_{2}$CrO$_{2}$Cr$_{2}$OAs$_{2}$ as an altermagnet that is highly anisotropic in its $J_2$  exchange couplings with large Néel temperature ($\sim 600$ K).
Using exchange coupling parameters extracted from DFT calculations,
we compute the magnon spectra for altermagnetic systems. 
As expected, chiral splittings in the magnon dispersion are directly
correlated with anisotropy between crystallographically inequivalent
$J_{2}$ exchange interactions.

\end{abstract}
\maketitle

\section{Introduction}
The Lieb lattice, a square lattice containing three inequivalent sublattices, was originally introduced by Lieb \cite{lieb_two_1989}
to describe electronic interactions in cuprate superconductors \cite{bednorz_possible_1986}.
In recent years, this lattice geometry has experienced a resurgence
of interest in the context of altermagnetism (AM) \cite{PhysRevLett.134.096703,kaushal_altermagnetism_2024,durrnagel_altermagnetic_2024}
--- a novel class of collinear magnets that exhibit alternating spin
polarization on symmetry-related sites, leading to distinctive transport
and topological phenomena despite vanishing net magnetization \cite{mazin_editorial_2022,smejkal_beyond_2022,mazin_prediction_2021,smejkal_anomalous_2022,smejkal_giant_2022,smejkal_emerging_2022}.

Interestingly, the earliest use of the term Lieb lattice referred
to the two-dimensional counterpart of the perovskite structure \cite{weeks_topological_2010}
--- essentially a layered, 2D perovskite-type lattice [see Fig.
\ref{fig:structure} (b)] --- which contrasts with the modern usage of the Lieb lattice
in the context of AM [see Fig. \ref{fig:structure}(d)], which is a 2D {\it anti}perovskite lattice. 
Therefore, to avoid the confusion, we will refer to this 2D antiperovskite lattice as inverse Lieb lattice (ILL)

In 2023, this two-dimensional antiperovskite square lattice was proposed
as a minimal toy model for exploring altermagnetic behavior inspired
by the cuprate (2D perovskite) systems \cite{brekke_two-dimensional_2023},
although this was done without awareness of the fact that real two-dimensional
antiperovskite structures already exist in nature. Much like their
three-dimensional counterparts, antiperovskite materials with the
general structure shown in Fig. \ref{fig:structure}(c), these 2D antiperovskites, in fact constitute
a diverse and largely untapped materials class, offering significant
promise for discovering new quantum phases and functional magnetic
properties \cite{shengComputationalApplicationsDiscovery2024}.

Many of these materials are metallic, making them ideal for experimental
studies, and some exhibit rather high N\`{e}el temperatures.

Furthermore, a particularly exciting aspect is the prevalence of $d$-wave
altermagnetism in these systems, a rare characteristic given that
most currently studied altermagnets (e.g., MnTe, CrSb) \cite{smejkal_emerging_2022}
are hexagonal and $g$-wave. This $d$-wave nature previously spurred
significant interest in, though it eventually
proved non-magnetic \cite{kesler_absence_2024,hiraishi_nonmagnetic_2024,smolyanyuk_fragility_2024}.

ILL offers a promising avenue for discovering new $d$-wave
altermagnetic materials. This family of materials is far more extensive
than commonly realized \cite{kabbour_structure_2008,free_synthesis_2011,he_synthesis_2011,sheath_structures_2022,song_tetragonal_2022,xiao_thcr2_2024,fuwa_crystal_2010,ni_physical_2010,free_low-temperature_2010,ablimit_v2_2018,ablimit_weak_2018},
and not all members exhibit altermagnetic properties. Therefore, understanding
the factors that control their ground states is paramount. Moreover,
this platform is ideally suited for exploring chiral magnons. Unlike
other systems where chirality arises from distant interactions --- such as the
10th and 11th nearest-neighbors (NNs) in MnTe \cite{liu_chiral_2024} or the seventh NN in rutile
\cite{gohlke_spurious_2023} --- the ILL allows for chirality
generation through significantly closer, second NN interactions,
which is a significant advantage for fundamental research and potential
applications \cite{songAltermagnetsNewClass2025}.

The paper is organized as follows. In Section \ref{sec:structure}, we introduce the general structure and definition of the ILL. 
Section \ref{sec:method} describes the computational details of our (density functional theory) DFT calculations and the methodology used to extract exchange coupling parameters. 
In Section \ref{sec:results} we begin our analysis by constructing phase diagrams based on a simple Heisenberg model to elucidate the fundamental mechanisms underlying altermagnetism and other complex magnetic phases observed experimentally. 
To bridge theory with experiment, we then investigate a series of known ILL compounds using DFT. 
By combining the extracted exchange parameters with the phase diagram analysis, we determine their magnetic ground states and compare our findings with available experimental results. 
For \ch{Sr2CrO2Cr2OAs2}, due to its unique crystal structure, as well as its distinctive electronic and magnetic properties, a dedicated subsection is provided for the compound.
Finally, we present the calculated magnon spectra for several selected altermagnetic compounds to visualize the impact of $J_2$ anisotropy on chiral magnon band splitting.
Through the paper, we use the following convention for the exchange Hamiltonian: $E=\sum_{\rm bonds}J_{ij}\mathbf{m}_i\cdot\mathbf{m}_i$, where summations is over all bonds (no double counting and $\mathbf{m}_i, \mathbf{m}_j$ are the unit vectors at the and of the bond parallel to local magnetic moments.

\section{Crystal structure}\label{sec:structure}

\begin{figure}
\centerline{\includegraphics[scale=0.4]{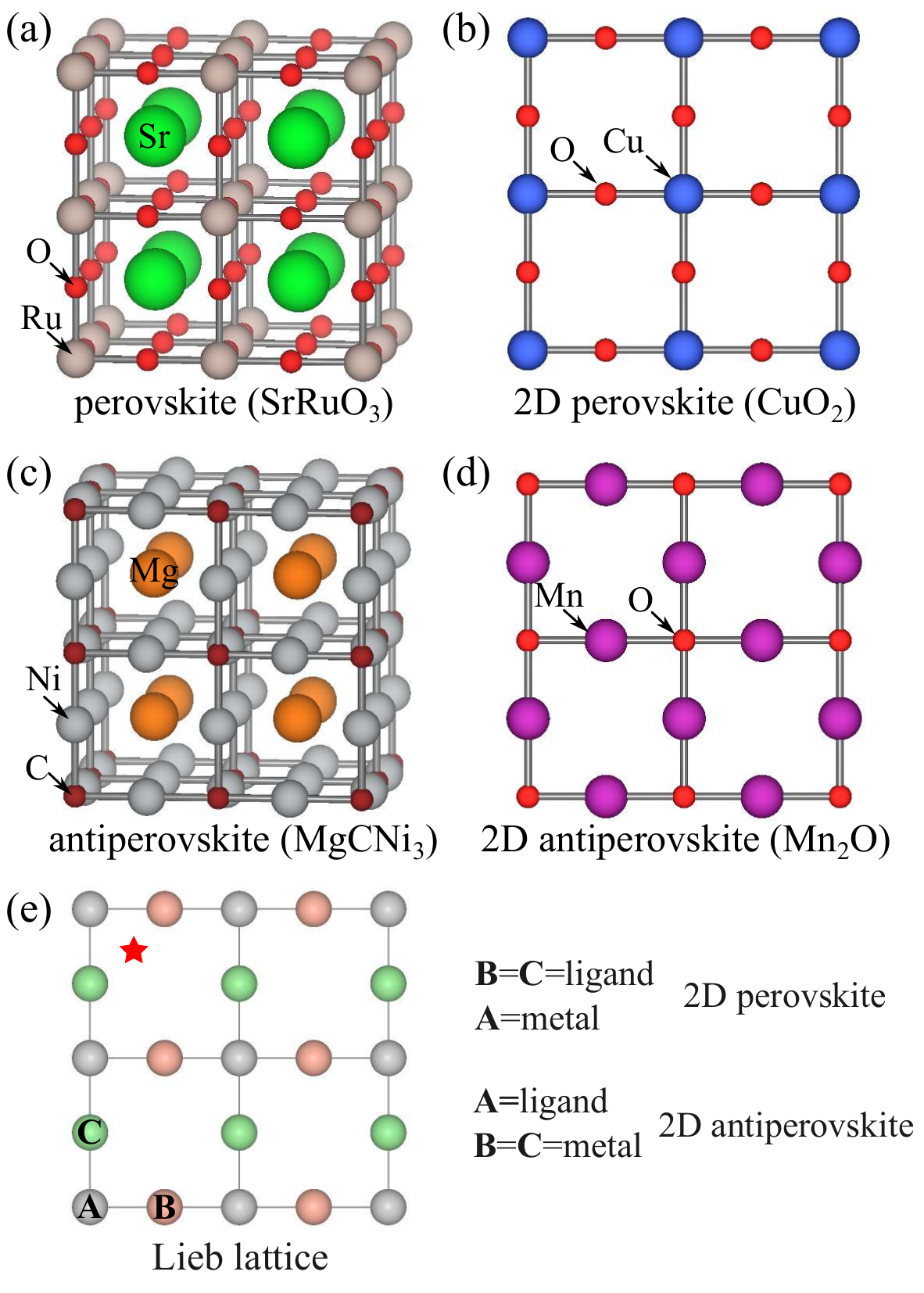}}

\caption{ Examples of 3D and 2D perovskite and anti-perovskite structures. The star shows the midpoint between the B and C sites, illustrating the absence of an inversion center at this point,}\label{fig:structure}
\end{figure}

\begin{figure}
\includegraphics[scale=0.45]{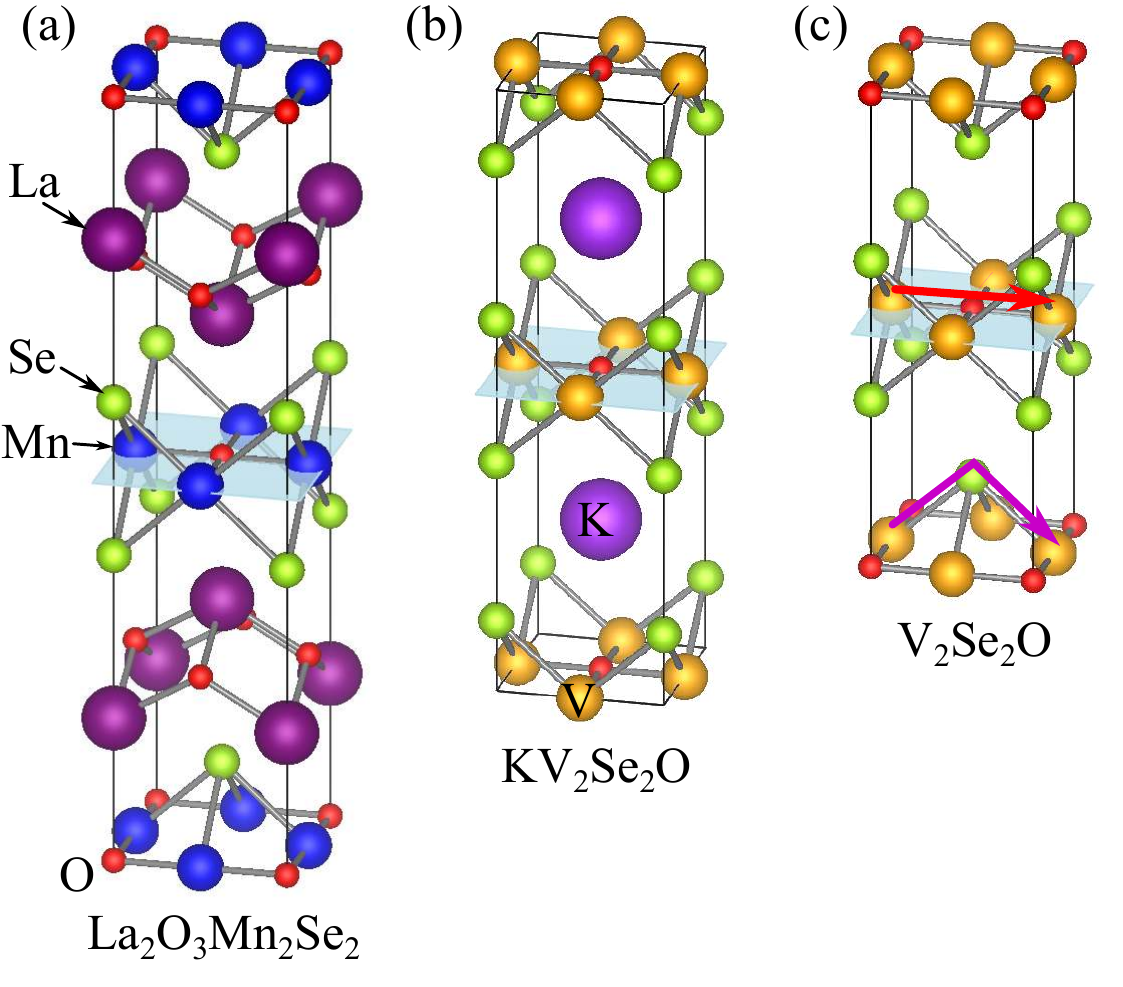}

\caption{Examples of different compounds that contain ILL structure
which are indicated by the light blue plane. Blue and red arrowed lines represent two inequivalent exchange pathways between second neighbors.}
\label{fig:compounds}
\end{figure}

Figure \ref{fig:structure} (e) illustrates the origin of altermagnetism in ILL. The symmetry condition for AM is that there is a symmetry operation that connects the spin-up and spin-down sublattices, and that this operation is neither translation nor inversion. In the ILL lattice, this operation is a (110) mirror, while inversion is present, but only maps each sublattice upon itself.

The ILL structure is abundant in nature, appearing in a variety
of transition metal compounds. Figure \ref{fig:compounds}  highlights
three representative examples containing ILL planes, typically
composed of transition metal (TM), oxygen, and commonly
chalcogen (e.g., Se) atoms, with or without intervening filler layers
between the ILL planes. 

A key structural feature of these systems is the presence of two distinct
second NN $J_{2}$ exchange pathways, as illustrated
in Fig. \ref{fig:compounds} (c): one coupling through an oxygen atom
with a 180° bond angle (red), and the other via a chalcogen atom, forming
approximately 90° bond angles (purple). This intrinsic anisotropy in exchange
pathways due to crystal symmetry plays a central role in shaping the
magnetic interactions and, in many cases, underpins altermagnetism
in these materials, a topic discussed in detail in the later sections. 

Moreover, the nature of the transition metal ion---likely through
variations in band filling---and the local crystal field environment
significantly influence the exchange interaction profile. For instance,
in the La$_{2}$O$_{3}$M$_{2}$Se$_{2}$ family (M=Mn, Fe or Co),
the Mn-based compound exhibit altermagnetism, while Co- and Fe-based
analogues do not \cite{wei_2_2025,fuwa_orthogonal_2010,gunther_magnetic_2014}.
Additionally, the presence of additional filler layers can modify
the crystal field symmetry and affect charge transfer processes, thereby
tuning both the metallicity and magnetic exchange couplings. 

A case in point is the contrast between insulating V$_{2}$Se$_{2}$O
\cite{lin_structure_2018} and conducting KV$_{2}$Se$_{2}$O \cite{jiang_discovery_2025},
both of which are likely altermagnetic despite their different electronic
ground states. Collectively, this structural and chemical flexibility
makes ILL-based compounds an exceptionally versatile platform for
investigating and engineering altermagnetism.

\section{Computational details}\label{sec:method}

First-principles calculations were performed using the numerical-orbital-based
\cite{ozakiVariationallyOptimizedAtomic2003} density functional theory
(DFT) code OpenMX \cite{OpenMX}. In these calculations, the core
electrons were treated using norm-conserving pseudopotentials \cite{morrisonNonlocalHermitianNormconserving1993,vanderbiltSoftSelfconsistentPseudopotentials1990}.
The Perdew-Burke-Enzerhof (PBE) \cite{perdewGeneralizedGradientApproximation1996}
generalized gradient approximation was employed to describe exchange-correlation effects. In addition, for the gapped systems, a Hubbard
U correction \cite{liechtensteinDensityfunctionalTheoryStrong1995,dudarevElectronenergylossSpectraStructural1998}
is used to account for the strongly correlated $d$-states in TMs. 

The detailed parameters regarding lattice parameters for the crystal structure, basis sets, and Hubbard $U$ values are provided in the Supplementary Materials.

After achieving self-consistency for the charge densities, the exchange
coupling constants $J_{i}$ were computed perturbatively via the Green's
function (GF) method \cite{antropovExchangeInteractionsMagnets1997,katsnelsonFirstprinciplesCalculationsMagnetic2000}
as implemented in OpenMX 3.9 \cite{terasawaEfficientAlgorithmBased2019,omxgf2004}.
This approach allows for the direct evaluation of exchange interactions
between arbitrary pairs of magnetic sites from a single magnetic configuration,
irrespective of interionic distance.

Since the GF method is based on a linear response framework around
the magnetic ground state, it is generally well suited for linearized spin-wave
calculations. For Sr$_{2}$CrO$_{2}$Cr$_{2}$OAs$_{2}$, however, the Green's function method yielded
exchange parameters that depend strongly on the reference magnetic configuration.
In this case, the exchange parameters were instead
determined using the standard Heisenberg model total-energy-mapping approach.

\section{Results and Discussion}\label{sec:results}

\subsection{Interactions and orders on the ILL}\label{sec:phases}

\begin{figure*}
\centerline{\includegraphics[scale=0.5]{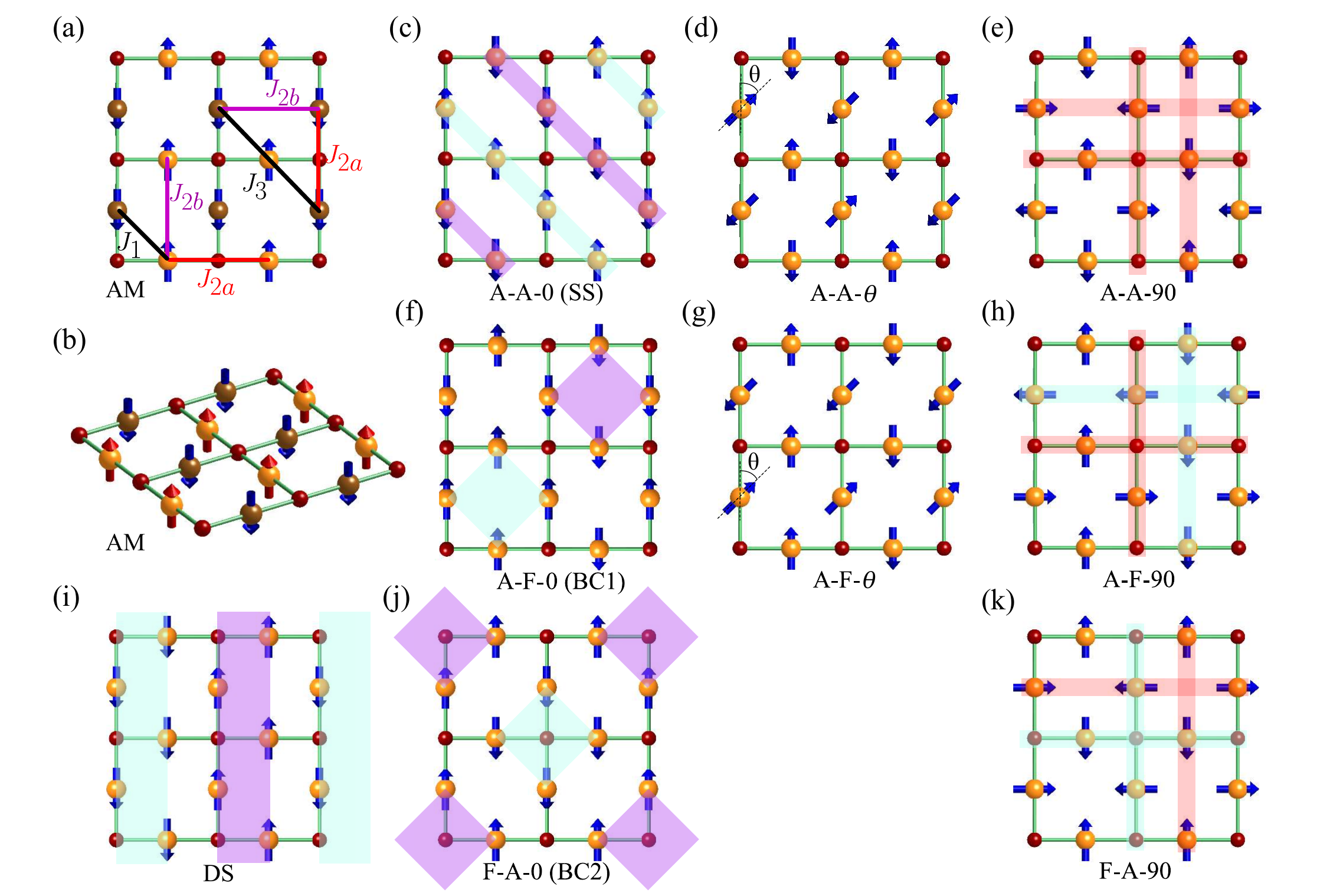}}

\caption{Definitions of the magnetic orderings on the simplest ILL plane
that contains oxygen (red) and transition metal (orange and brown)
sites. (a) and (b) are the altermagnetic states. (c), (d) and (e) correspond to the $\sqrt2\times\sqrt2$ order, and represent a continuous transition by rotating one spin sublattice with respect to the other, from the single-stripe (SS) to the ``A-A-90'' phase (in which the second neighbors are AF ordered in both directions (as shown by pink shades). (f), (g) and (h) show the effect of such rotation on a $2\times2$ order in which the spins form FM \pc{``blocks'' indicated by purple and light blue plaquettes} which are ordering with respect to each other in a checkerboard fashion (block-checkerboard, BC1 and BC2, structures). (i) is the double-stripe phase. (j) and (k) are similar to (f) and (h) except that now \pc{each FM plaquette formed around an Oxygen site}. Note that in the corresponding 90$^\circ$ structures the half of the second neighbors are ordered FM and half AF, as shown by the shades, and we call them A-F and F-A phases, respectively.
}\label{fig:config}
\end{figure*}

Aside from its connection to altermagnetism, the ILL lattice structure is by itself
interesting. The rich phase diagram characterized by both conventional
and highly degenerate magnetic orderings can occur depending on the
competition between distinct exchange interactions. Below we present
a detailed analysis of the phase diagram based on a simple $J_{1}$-$J_{2a}$-$J_{2b}$
model \cite{zhu_band_2010}, both with and without further NN $J_{3}$, where
$J_{2a}$ and $J_{2b}$ are two inequivalent second NN exchange interactions.
The exchange paths are defined in Fig. \ref{fig:config} (a).

According to the crystallographic symmetry, the ILL can be
viewed as comprising two sublattices. For better visualization, the
two sublattices of the same transition metal element are
coded in dark brown and orange colors as shown in Fig.
\ref{fig:config} (a) and (b) which are equivalent in the isotropic
Heisenberg model. The sites connected by either of the $J_{2}$ bonds
belong to the same sublattice, while different sublattices are connected
by $J_{1}$ bonds. In other words, for any given site, all four first/second
nearest neighbors are from different/same sublattices. 

\begin{figure}
\includegraphics[scale=0.38]{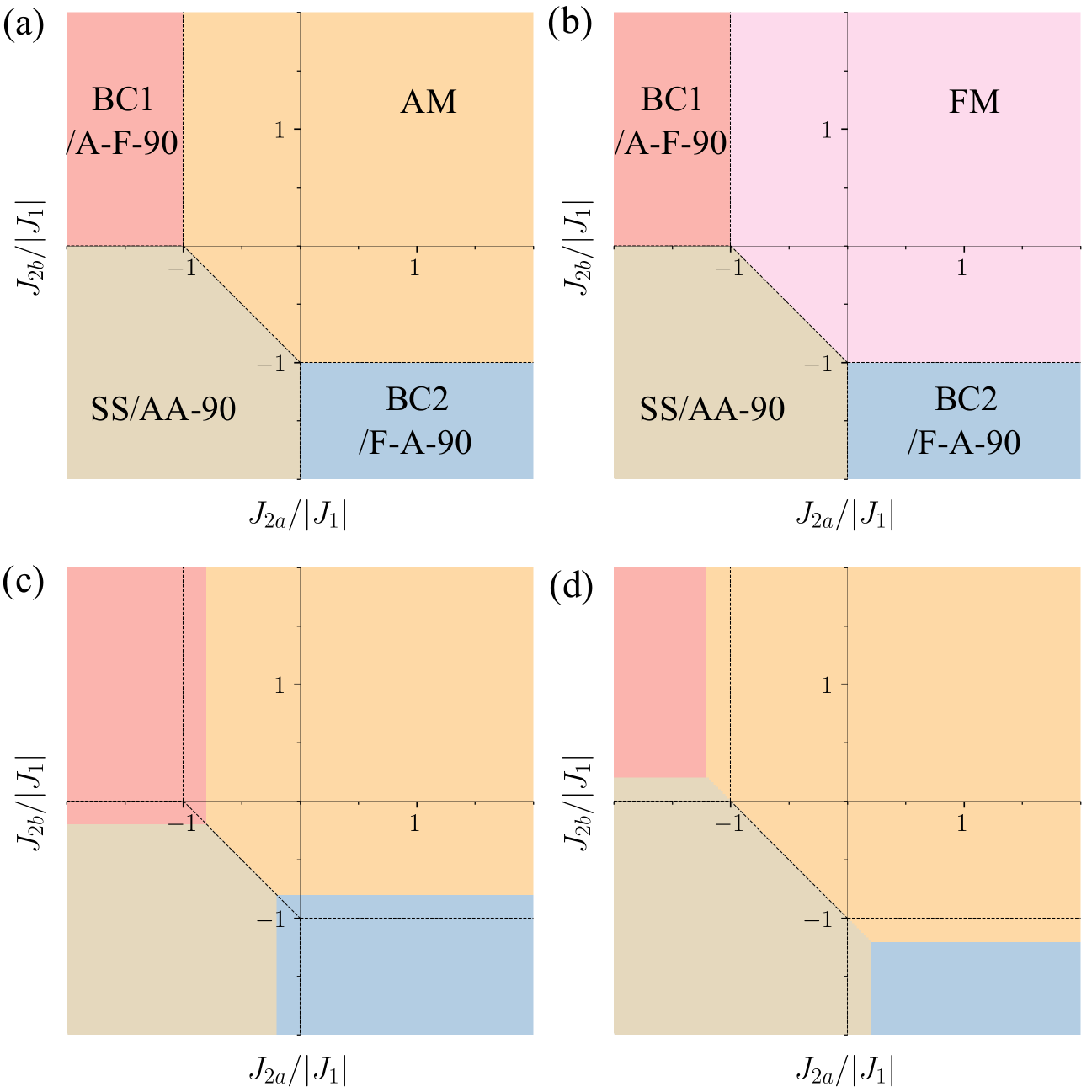}
\caption{Phase diagrams: (a), (c) and (d) for $J_{1}<0$ with $J_{3}=-0$,
$0.1|J_{1}|$and $0.1|J_{1}|$ respectively and (b) for $J_{1}>0$
with $J_{3}=0$. The blue lines in (c) and (d) are the phase boundaries
for $J_{3}=0$.}\label{fig:phase}
\end{figure}

We begin by considering antiferromagnetic $J_{1}$ case where the
corresponding phase diagram is shown in Fig. \ref{fig:phase} (a).
When $J_{1}$ is the dominant antiferromagnetic interaction, the system
naturally adopts the altermagnetic (AM) ordering regardless of the
signs of $J_{2a}$ and $J_{2b}$, since all first nearest neighbors have the opposite
spins, as defined in Figs. \ref{fig:config} (a) and (b). The AM phase
remains stable under two primary conditions: either $J_{1}$ dominance
(i.e. $|J_{1}|>|J_{2a/b}|$ for antiferromagnetic $J_{2}'s$) or ferromagnetic
$J_{2}$ interactions of any strength, as demonstrated in the phase diagram, where
the entire positive (FM) $J_{2a}$-$J_{2b}$ quadrant exhibits AM
ordering. 


The crystallographic distinction between $J_{2a}$ and $J_{2b}$ interactions
has profound implications for magnetic ordering.  In addition, the
presence of at least one dominant antiferromagnetic $J_{2}$ creates
significant frustration which manifests as a complex competition between
different magnetic phases including,  
as illustrated in Figs. \ref{fig:config} (c-k), (f-h) and (j-k),
three sets of continuously degenerate states.
Note that additional colored stripes and square plaquettes are introduced to help identify
the patterns. Except for the red stripes in Fig. \ref{fig:config} (e), (h) and (j) which
represent AF chains, all the rest of the colored objects indicate FM clusters.

To systematically track the complex magnetic states, we also introduce a more descriptive naming convention for the states illustrated in Figs. \ref{fig:config} (c-h) and (j-k). Each state is labeled using the general expression l-m-$\theta$, where l and m represent the spin alignment through coupling parameters $J_{2a}$ and $J_{2b}$ respectively, with each taking values of either A (antiferromagnetic) or F (ferromagnetic) and $\theta$ indicates the rotation angle of one sublattice with respect to the other as shown in Figs. (d) and (g).

If nearest neighbors in each individual sublattice order antiferromagnetically, then this frustration leads to complete decoupling between the two sublattices
at the Heisenberg level. The exchange energy between sublattices becomes
zero and therefore independent of the rigid rotations of spins in one sublattice relative
to the other, resulting in a continuous ground state degeneracy. When
both $J_{2}$ are antiferromagnetic, the SS phase, depicted in Fig.
\ref{fig:config} (c), becomes degenerate with A-A-$90$ phase as well as the intermediate states of arbitrary angle $\theta$ shown in Fig.
\ref{fig:config} (d) and (e). 


Conversely, when $J_{2b}$ is ferromagnetic with $J_{2a}$ being dominant
and antiferromagnetic, the BC1 phase exhibits complete degeneracy
with the noncollinear A-F-$90^{\circ}$ phase, as observed experimentally in
compounds like La$_{2}$O$_{3}$Fe$_{2}$Se$_{2}$ \cite{gunther_magnetic_2014,mccabe_weak_2014,freelon_magnetic_2019}.

Similarly, reversing the signs of the two  $J_{2}$'s
(i.e. $J_{2a}<0$ and $J_{2b}>0$) is equivalent to interchanging the two sublattices. As a result, both BC1 and 
all its degenerate states (A-F-$\theta$) will transform
into a different set of degenerate phases: BC2 and F-A-$\theta$ as indicated
in \ref{fig:phase} (a) with their corresponding magnetic orderings
illustrated in Figs. \ref{fig:config} (j) and (k). In all three situations
where there is at least one dominant antiferromagnetic $J_{2}$ that
outweighs $J_{1}$, it is easy to verify that any given site in one
sublattice is always surrounded by an equal number of up and down spins. 
A straightforward justification is provided in the Supplementary Materials.
The BC phase, while not commonly observed, has been theoretically
discussed in the context of iron telluride \cite{plaquette} and some
ILL materials such as Na$_{2}$Fe$_{2}$Se$_{2}$O \cite{suetin_electronic_2012}.
Interestingly, in iron-based superconductors $J_{2a}=J_{2b}$ by symmetry, and the BC1 and BC2 become degenerate, and also degenerate with the often-discussed double stripe (DS) phase shown in Fig. \ref{fig:config} (i) experimentally observed in FeTe \cite{plaquette,totalenergyanalysis0}. In the ILL, however, DS can never have the lowest energy, and BC1 and BC2 become inequivalent and competing.

Adding third and further neighbors does not remove this continuous degeneracy,  and the actual magnetic ground state
is determined by secondary interactions typically much weaker than
the primary exchange terms. Similar to iron-based superconductors,
factors such as single-site magnetocrystalline anisotropy, anisotropic
exchange interactions and biquadratic coupling become decisive in
selecting the true ground state from the degenerate manifold \cite{Wysocki2011,Natasha,Rice}.

Interestingly, the same frustration that decouples the two sublattices
persists as long as (at least one) $J_{2}$ dominates, regardless
of the sign of $J_{1}$. One can observe that even if $J_{1}$ is
ferromagnetic, as shown in Fig. \ref{fig:phase}(b), only the originally
AM phase will be replaced by the ferromagnetic (FM) phase, while the
rest of the phase diagram remains unchanged.

In a metallic system, the interactions between further neighbors $J_{3}$
can be appreciable. However, they generally do not introduce more phases
but rather shift the phase boundaries. Figs. \ref{fig:phase}(c) and (d)
show the phase diagrams for $J_{3}/|J_{1}|=-0.1$ and 0.1. The black dashed
lines indicate the original phase boundaries in the absence of $J_{3}$. 

\subsection{Connection to real materials}

\begin{table}

\caption{
Calculated exchange parameters $J_i$ (in meV) and experimentally observed magnetic ground states in representative compounds characterized by different filler and TM and $d$-shell filling $N_{3d}$.
}


\fontsize{8}{8}\selectfont
\def\arraystretch{2.0}
\begin{tabular}{l|l|cccc|c}
\hline 
Compound & Ordering & $J_{1}$ & $J_{2a}$ & $J_{2b}$ & $J_{3}$ & $N_{3d}$\tabularnewline
\hline 
La$_{2}$O$_{3}$Mn$_{2}$Se$_{2}$ & AM \cite{wei_2_2025} & -31.6 & -7.5 & -6.1 & 0.2 & $d^{5}$\tabularnewline
La$_{2}$O$_{3}$Co$_2$Se$_{2}$ & A-A-$90$ \cite{fuwa_orthogonal_2010} & -20.1 & -50.6 & 1.2 & 0.1 & $d^{7}$\tabularnewline
La$_{2}$O$_{3}$Fe$_{2}$Se$_{2}$ & A-F-90 \cite{gunther_magnetic_2014,mccabe_weak_2014} & -27.1 & -51.0 & 7.2 & -0.9 & $d^{6}$\tabularnewline
Na$_{2}$Fe$_{2}$Se$_{2}$O & BC \cite{he_synthesis_2011} & -13.1 & -51.5 & 3.2 & -0.6 & $d^{6}$\tabularnewline
V$_{2}$Se$_{2}$O &AM\footnote{There is no experimental data for the magnetic ground state.} \cite{yu_high_2021,ma_multifunctional_2021} & -24.7 & -2.6 & 8.1 & 0.6 & $d^{2}$\tabularnewline
KV$_{2}$Se$_{2}$O &AM \cite{jiang_discovery_2025} & -41.6 & 34.1 & 0.8 & 7.0 & $d^{2.5}$\tabularnewline
RbV$_{2}$Te$_{2}$O &AM \cite{zhang_crystal-symmetry-paired_2025} & -38.6 & 31.8 & -5.3 & 7.1 & $d^{2.5}$\tabularnewline
Sr$_{2}$CrO$_{2}$Cr$_{2}$OAs$_{2}$ &AM \cite{sheath_structures_2022} & -66.7 & -19.1 & 34.4 & -8.5 &  $d^{3+x}$\tabularnewline
\hline 
\end{tabular} \label{tab:jex}
\end{table}

\begin{figure}
\centerline{\includegraphics[scale=0.6]{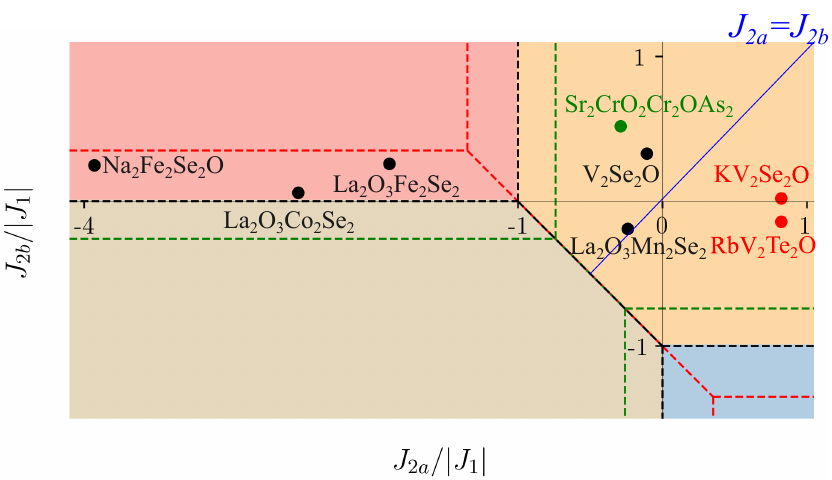}}
\caption{Phase diagram with the calculated data points for the compounds listed in TABLE \ref{tab:jex} The colored dashed lines represent the phase boundaries corresponding to data points of the same color. The blue dashed line within AM phase, which indicates the condition where $J_{2a}=J_{2b}$.}\label{fig:phase2}
\end{figure}

Although the ILL offers a promising avenue for discovering
new $d$-wave altermagnetic materials, 
not all members of the family are altermagnetic. Table
\ref{tab:jex} lists the calculated exchange coupling parameters
up to fourth NN, the resulting magnetic ground state, and the $d$-shell
filling for a series of representative compounds characterized by different filler layers and TMs.
The exchange coupling data of up to third NN are also visualized in Fig. \ref{fig:phase2} within a phase diagram.
The colored dashed lines represent the phase boundaries corresponding to data points of the same color and the solid blue line within the AM phase indicates the condition where $J_{2a}=J_{2b}$. 
Since two compounds \ch{KV2Se2O} and \ch{RbV2Te2O} have very close $J_3/|J_1|$ ratios, $0.17$ and $0.18$ respectively, the phase boundary for the two is estimated with the average value $0.175$.
To observe a substantial chiral magnon splitting, significant anisotropic $J_2$ coupling is required within the AM phase.
When the condition $J_{2a}=J_{2b}$ is satisfied, the $J_2$  interaction becomes isotropic, resulting in the absence of magnon splitting.

Despite sharing the same ILL structure, the details of the
exchange interactions and, as a result, the magnetic ordering, depend
strongly on the TM and filler layers. For example, the first three compounds
in Table \ref{tab:jex} belong to the same La$_{2}$O$_{3}$M$_{2}$Se$_{2}$
family with the transition metal M (Mn, Co, or Fe) being the only difference. However, the qualitative behavior can vary quite drastically.

La$_{2}$O$_{3}$Mn$_{2}$Se$_{2}$ exhibits a dominant $J_{1}$ interaction
with much weaker $J_{2a}/J_{2b}$ interactions, which, according to the phase diagram,
results in a stable AM phase \cite{wei_2_2025}. Interestingly, despite their crystallographic inequivalence, the
$J_{2}$ interactions are almost isotropic in this compound, with $J_{2a}\approx J_{2b}$ \cite{wei_2_2025}.

On the other hand, La$_{2}$O$_{3}$Co$_{2}$Se$_{2}$ and La$_{2}$O$_{3}$Fe$_{2}$Se$_{2}$
have dominant antiferromagnetic $J_{2a}$ interactions,
and both were reported to adopt noncollinear magnetic phases: A-A-90 for M=Co \cite{fuwa_orthogonal_2010,free_synthesis_2011}
and A-F-90 for M=Fe \cite{gunther_magnetic_2014,mccabe_weak_2014,freelon_magnetic_2019}.
These magnetic orderings, depicted in Figs. \ref{fig:config} (e) and (h),
are degenerate with the respective
collinear SS [Fig. \ref{fig:config}(c)] and BC1 [Fig. \ref{fig:config}(f)] phases at the Heisenberg level, as discussed in Section \ref{sec:phases}. The noncollinearity
in both cases is likely driven by biquadratic interactions and single-ion anisotropy effects.

Although the Co and Fe compounds exhibit similar exchange profiles according to our phase diagram analysis, they are expected to display the same magnetic ground state. However, despite correctly predicting both compounds to be non-altermagnetic, the theoretical model fails to reproduce the distinct magnetic states observed experimentally. As illustrated in Fig. \ref{fig:phase2}, \ch{La2O3Co2Se2} has a smaller $J_{2b}$ and lies very close to the SS/BC1 phase boundary, indicating that its SS phase is less stable than that of \ch{La2O3Fe2Se2}. This suggests that the true ground state of \ch{La2O3Co2Se2} likely involves additional interactions beyond the Heisenberg model.

A closely related Na$_{2}$Fe$_{2}$Se$_{2}$O compound \cite{he_synthesis_2011}, which
has the same {[}Fe$_{2}$O{]}$^{2+}$ ILL sheets but with the fluorite-like oxide
layers replaced by layers of Na$^{+}$ ions, exhibits
a dominant $J_{2a}$ interaction (mediated through oxygen) similar
to the above Co and Fe compounds. Its magnetic ordering 
is not known experimentally, and was theoretically suggested to be one of the 
blocked-checkerboard magnetic patterns \cite{suetin_electronic_2012}.

The next three on the list are vanadium-based compounds. 
Both KV$_{2}$Se$_{2}$O \cite{jiangMetallicRoomtemperatureDwave2025} and RbV$_{2}$Te$_{2}$O
\cite{zhang_crystal-symmetry-paired_2025} have been determined 
to be altermagnetic both theoretically and experimentally 
with N\'{e}el temperatures around room temperature. 
These findings are further corroborated by our calculated exchange parameters.

While these compounds also exhibit stable AM phases,  their exchange interaction profiles differ significantly from La$_{2}$O$_{3}$Mn$_{2}$Se$_{2}$.
The major distinction lies in their large ferromagnetic $J_{2a}$, which is comparable to $J_{1}$ in magnitude and further stabilizes the AM ordering, explaining the large observed $T_{N}$. 

Strong FM interaction for a $180^{\circ}$ superexchange is inconsistent with the Hubbard model and indicates that metallicity of the partially-filled $t_{2g}$ bands plays an important role in these compounds. This metallicity also enhances longer-range $J_3$ interactions.

A pure V$_{2}$Se$_{2}$O bulk system without any
intervening filler layers has also been successfully synthesized \cite{lin_structure_2018}. To the best of our knowledge,
no experimental data on its magnetic ground state have been reported.
However, several theoretical studies, including our own DFT calculations
included in Table \ref{tab:jex}, suggest that this compound is likely
altermagnetic \textcolor{red}{\cite{yu_high_2021,ma_multifunctional_2021}}. 

Unlike the above-mentioned two metallic vanadium-based
compounds, V$_2$Se$_2$O 
appears to be an insulator \cite{lin_structure_2018}, probably thanks to integer occupancy and considerable tetragonal crystal-field splitting of the $t_{2g}$ band. Consistent with its insulating
nature, the range of the exchange interaction becomes much shorter,
and $J_{2a}$, which would need to be strongly FM for stabilizing AM, is not even FM at all anymore.
This contrast serves as a good demonstration
on how the $d$-band occupation affects the magnetic and electronic properties.

The last entry in Table \ref{tab:jex} is Sr$_{2}$CrO$_{2}$Cr$_{2}$OAs$_{2}$, which
presents a particularly interesting case with mixed characteristics.
Although it has not been previously discussed in the context of altermagnetism,
an earlier experimental study has determined its magnetic ordering that we can now recognize as
 altermagnetic \cite{sheath_structures_2022}.

Our calculated results are consistent with the experimental magnetic structure,
suggesting that the AM ordering is the magnetic ground state stabilized
by the dominant $J_{1}$ interaction. More notably, this compound
exhibits large anisotropy in $J_{2}$ interactions unlike other cases
in this study: both $J_{2a}$ and $J_{2b}$ are now large in magnitude and opposite in sign. 

A particularly interesting twist is that this compound forms a natural heterostructure of the original Lieb-model (cuprate-like) CrO$_2$ planes and AM ILL Cr$_2$O planes. By symmetry, there is no Heisenberg coupling between the two. The original paper \cite{sheath_structures_2022} suggested full charge disproportionation between the two Cr site, the one in the AM plane being Cr$^{3+}$ ($d^3$) and the other one Cr$^{2+}$ ($d^4$). However, the material being metallic, full disproportionation is unlikely, and our calculations suggest occupations of $d^{3+x}$ and $d^{4-x}$, with $x\sim0.2$--0.3. Again, metallicity plays an important role, moving  $J_2$ interactions toward FM and boosting longer-range coupling.

Table \ref{tab:jex} strongly suggests that the $d$-band filling plays a crucial role in shaping the magnetic interactions in ILL and eventually generating AM. Indeed, while all studied materials have a negative $J_1$ conducive to AM, it is particularly large for $d^5$ and $d^{3\pm x}$ occupations. Further, six or more electrons in the $d$-shell are associated with strong negative $J_2$, which is very unfavorable for AM. Fractional occupation and the resulting metallicity favor a FM component in $J_2$, which tends to stabilize the AM state and increase its ordering temperature. We have not been able to find a simple microscopic explanation for all these trends, so they remain empirical --- but very helpful in the search of new high-temperature metallic ILL AM materials.

\subsection{A special case of \ch{Sr2CrO2Cr2OAs2}}

\begin{figure}[htb]
    \centerline{\includegraphics[scale=0.37]{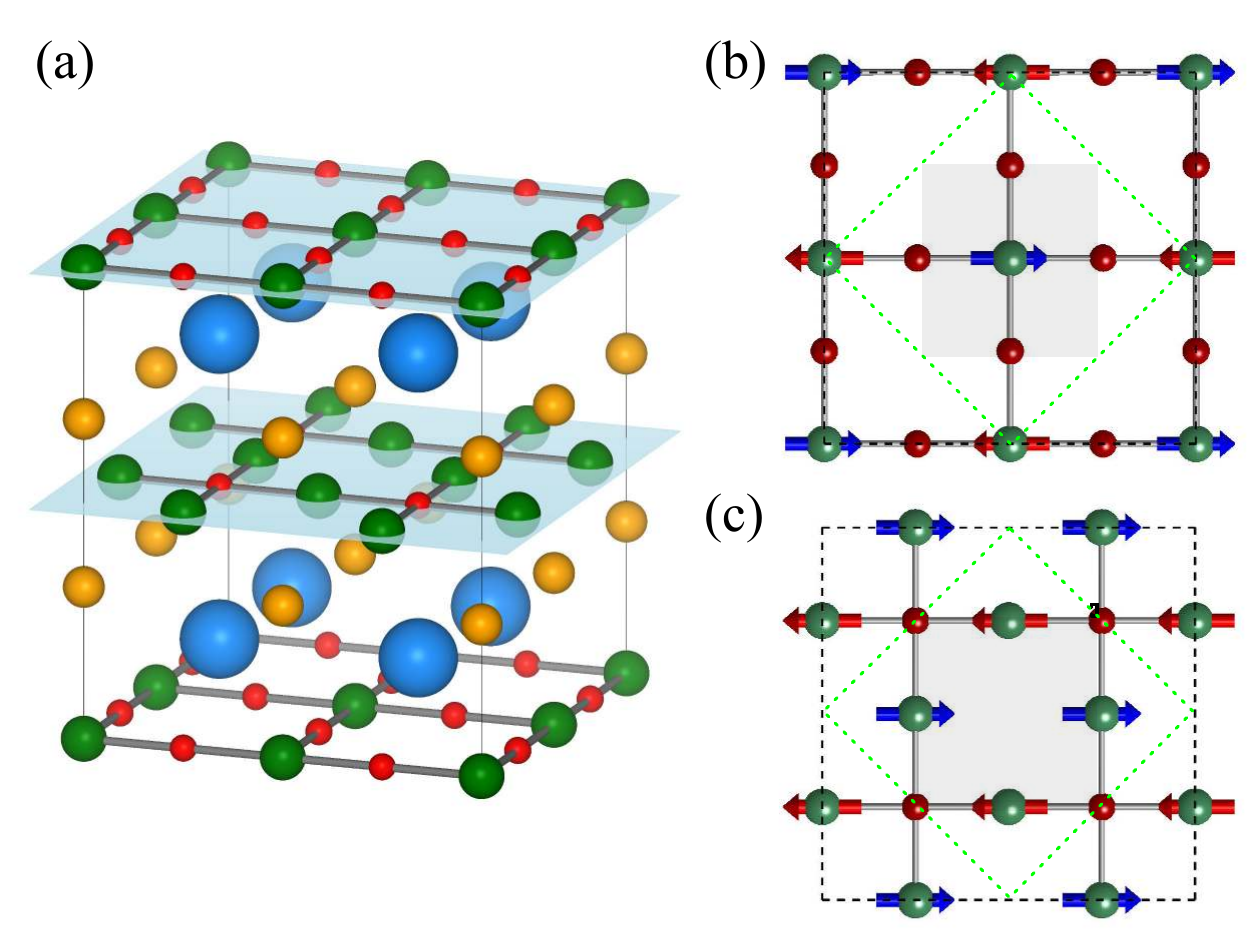}}
    \caption{Crystal structure of Sr$_{2}$CrO$_{2}$Cr$_{2}$OAs$_{2}$:
    (a) full supercell, (b) 2D perovskite plane, and (c) antiperovskite
    (ILL) plane. Black dashed lines, green dotted lines, and light-gray shading in panels (b) and (c) show, respectively, the $2\times2$ computational cell, the $\sqrt{2}\times\sqrt{2}$ primitive unit cell for the magnetic ground state, and the crystallographic primitive unit cell.}\label{fig:As_struct}
\end{figure}

Among the series of compounds we investigated, \ch{Sr2CrO2Cr2OAs2}
emerges as the most compelling system and deserves dedicated discussion.
As seen in Fig. \ref{fig:As_struct}(a), this compound exhibits a unique structural architecture containing both perovskite
and antiperovskite (ILL) planes with chromium atoms in different
oxidation states: \ch{Cr^2+} in the perovskite  and \ch{Cr^3+} in the ILL plane, respectively.

An experimental study has established that the ILL plane is ordered altermagnetically, while the perovskite plane adopts a checkerboard
antiferromagnetic ordering depicted
in Fig. \ref{fig:As_struct}(b) and \ref{fig:As_struct}(c), respectively \cite{sheath_structures_2022}. Our calculated
exchange parameters for the ILL plane (Table \ref{tab:jex}) strongly support the
observed altermagnetic ordering in that plane. The exchange coupling in the perovskite plane ($J_1=-57.9$, $J_2=-2.2$ meV) is dominated by antiferromagnetic nearest-neighbor coupling, which stabilizes checkerboard antiferromagnetism. Thus, magnetic ordering in both types of planes is stabilized by strong antiferromagnetic nearest-neighbor coupling. These large interactions explain the remarkably
high N{\'e}el temperature of approximately $600$ K \cite{sheath_structures_2022}.

The different symmetries between the magnetic orderings in two types of
lattice planes introduce additional  complexity. 
Antiferromagnetic ordering in the perovskite layers lowers the symmetry and results in a 
$\sqrt{2}\times\sqrt{2}$ doubled magnetic unit cell indicated by the green dotted line in Fig. \ref{fig:As_struct}(b) and \ref{fig:As_struct}(c),
where the crystollographic primitive cell is shown by gray shading. The interlayer interactions are considerably weaker
than the in-plane parameters, and they are inherently frustrated by the magnetic ordering, regardless of the stacking configuration.

Thus, Sr$_{2}$CrO$_{2}$Cr$_{2}$OAs$_{2}$ is a unique case when AM, generally believed to be associated with a $q=0$ magnetic order, now forms a structure with two $q$-vectors, one of which is non-zero.
The possibility of AM involving nonzero $q$-vectors has been pointed out before \cite{Libornpb}, but the examples given there are rather complicated and nontrivial, while this natural heterostructure is a very clean and simple example of a ``supercell AM,'' using the term introduced in Ref. \cite{Libornpb}. 
In the present case, it is clear that magnetic unit cell doubling does not destroy AM because the finite $q$-vector is not associated with the altermagnetic ILL plane. In other words, simply adding another piece to the magnetic structure can only \emph{lower} the magnetic symmetry of the whole system. The only question is whether the equivalence between the magnetic sublattices is retained, as it is in the present case.

It is interesting that the addition of the checkerboard-ordered perovskite CrO$_2$ layers does not even lower the spin point group compared to the standalone altermagnetic ILL. The latter is ${}^24/{}^1m_z{}^2m_d{}^1m_y$ in our crystallographic setting. The ${}^1m_z$ and ${}^1m_y$ generators are obviously not disturbed by the checkerboard CrO$_2$ layers. There are two types of simple ${}^24$ axes passing through either O or As atoms; the latter project onto the center of the shaded area in Fig. \ref{fig:As_struct}(c). There are also two types of [110] mirror planes: one simple ${}^2m_d$ passing through O and As atoms, and another \emph{glide} ${}^2n_d$ mirror plane passing through the Cr atoms. The addition of the CrO$_2$ layers breaks only the ${}^24$ axes passing through As atoms but is compatible with those passing through the O atoms of the ILL. It also breaks the simple ${}^2m_d$ mirror planes but is compatible with the glide ${}^2n_d$ mirror planes. Thus, Sr$_{2}$CrO$_{2}$Cr$_{2}$OAs$_{2}$ has the same AM spin \emph{point} group ${}^24/{}^1m_z{}^2m_d{}^1m_y$ as a standalone AM ILL.

Within the ILL plane, beyond the dominant $J_{1}$ interaction that
stabilizes the altermagnetic ordering, the second NN
exchange interactions exhibit remarkable anisotropy. In contrast to
other systems in our study, $J_{2b}$, as mentioned, is not only substantial but also opposite in sign to $J_{2a}$. This strong anisotropy, as discussed below, generates enormous chiral magnon splitting of $\approx 60$\% of the total magnon band width --- an amplitude so far unheard of. 

\begin{figure}[htb]
    \centerline{\includegraphics[scale=0.56]{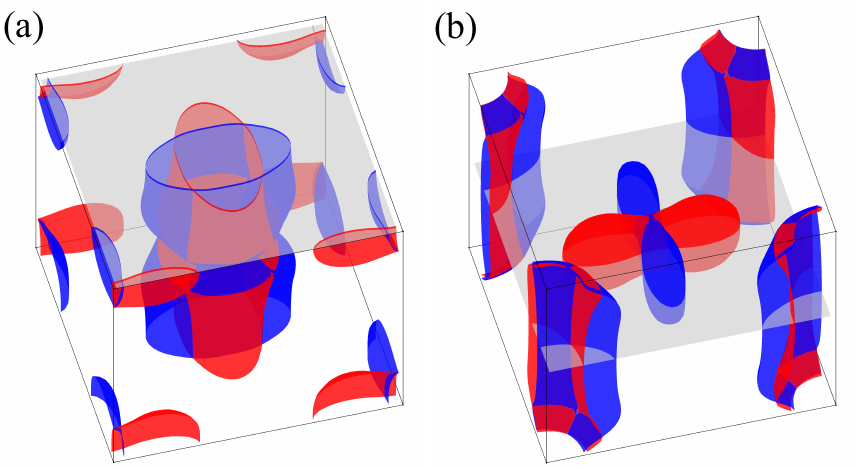}}
    \caption{Fermi surfaces in the first BZ of Sr$_{2}$CrO$_{2}$Cr$_{2}$OAs$_{2}$ which is centered (a) around the $\Gamma$ point, or (b) around its corner.}
    \label{fig:fermi_srf}
\end{figure}

The strong anisotropy in $J_{2}$ exchange interactions is consistent with a uniquely strong anisotropy
in the electronic band structure. Figure \ref{fig:fermi_srf} shows
the same Fermi surface in two representations: centered at the $\Gamma$-point [panel (a)] and at the corner of the Brillouin zone (BZ) [panel (b)]. Two distinct groups of Fermi surface pockets exhibit
spin splittings, with highly anisotropic Fermi surfaces from the two spin
channels related by a 90-degree rotation. The gray planes at $k_z=0.5\pi/c$ indicate where
the largest splitting occurs for both groups of bands.

The first set of Fermi pockets is centered around $(k_{x},k_{y})=(0,0)$ and extends
along the $k_{z}$ direction. Near the $\Gamma$-point, both pockets
appear relatively isotropic and nearly degenerate, becoming increasingly
anisotropic as they move away from the $k_{z}=0$ plane.

The second group of Fermi pockets exhibits substantial splitting near the BZ corners.
The cross-section projection onto the $k_{z}=0.5\pi/c$ plane resembles
a pair of propeller blades intersecting at $90$ degrees,
with exceptionally large momentum separation between  the spin-split Fermi surfaces, as shown in Fig. \ref{fig:fermi_srf}
(b).

The Sr$_{2}$CrO$_{2}$Cr$_{2}$OAs$_{2}$ system represents a fascinating
example.  The coexistence of altermagnetic and antiferromagnetic layers, a gigantic chiral magnon splitting, an
extremely high Néel temperature, and ultra-strong spin anisotropy of the Fermi surface position this compound as an excellent candidate for further theoretical and experimental investigations, as well as for device applications.

\subsection{Magnon spectra}

\begin{figure}
\centerline{\includegraphics[scale=0.37]{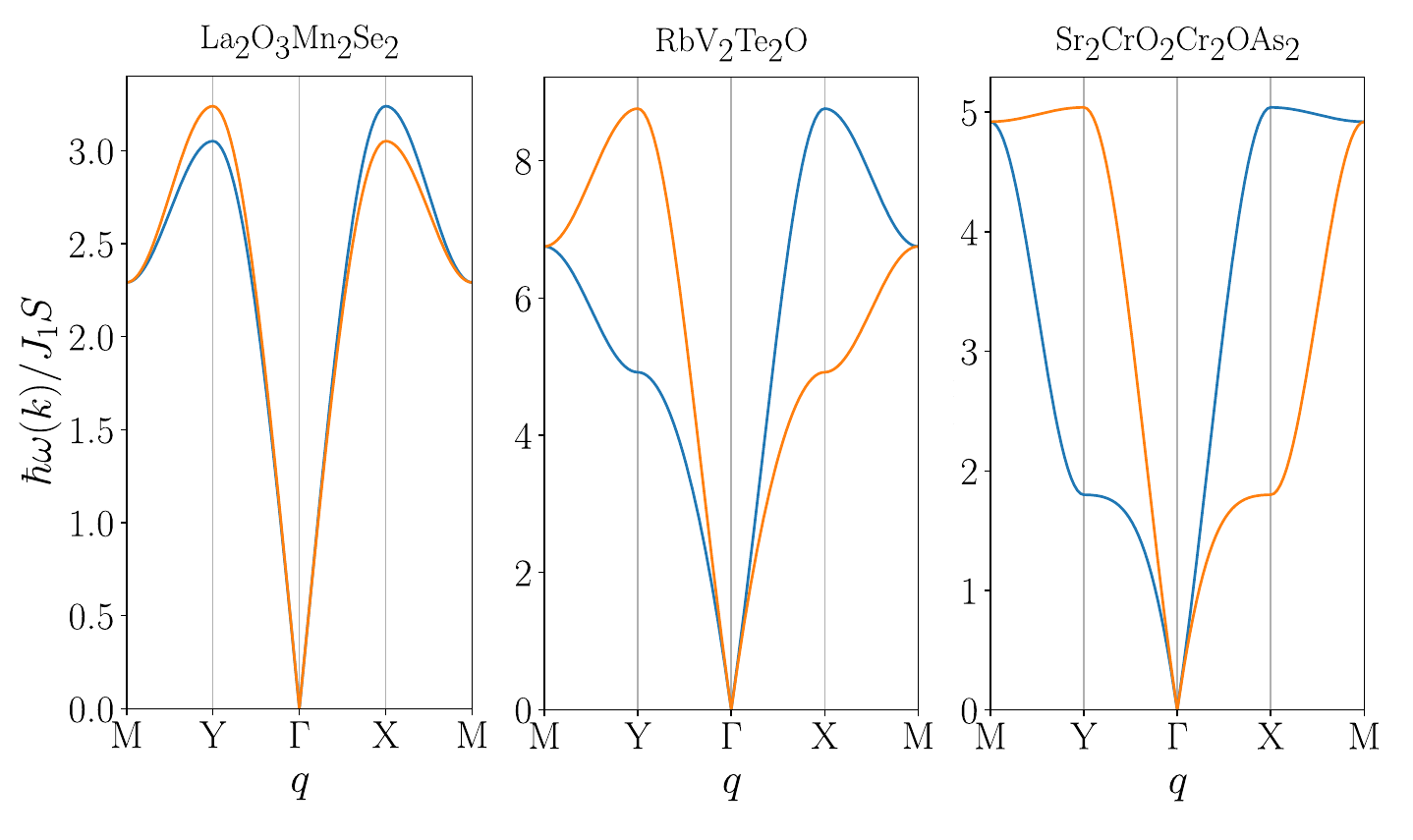}}

\caption{ Magnon spectra along high symmetry directions for (a) La$_{2}$O$_{3}$Mn$_{2}$Se$_{2}$,
(b) RbV$_{2}$Te$_{2}$O, and (c) Sr$_{2}$CrO$_{2}$Cr$_{2}$OAs$_{2}$ where we only consider magnons in the ILL layer.}\label{fig:magnon}
\end{figure}

Figure \ref{fig:magnon} shows the magnon spectra calculated using
the exchange coupling parameters listed in Table \ref{tab:jex} for
La$_{2}$O$_{3}$Mn$_{2}$Se$_{2}$,
RbV$_{2}$Te$_{2}$O, and Sr$_{2}$CrO$_{2}$Cr$_{2}$OAs$_{2}$, which, according to Table \ref{tab:jex}, have  
very different exchange parameter profiles.
As discussed in Section. \Ref{sec:phases}, the magnitude of the chiral magnon
splitting is controlled by the difference between $J_{2a}$ and $J_{2b}$.

In La$_{2}$O$_{3}$Mn$_{2}$Se$_{2}$, consistent with the earlier
theoretical work \cite{garcia-gassullMicroscopicOriginMagnetic2025}, the nearly isotopic $J_{2}$ values result in only
subtle splittings. In contrast, RbV$_{2}$Te$_{2}$O exhibits much
larger splitting thanks to the strong anisotropy of $J_{2}$ couplings.

In the case of Sr$_{2}$CrO$_{2}$Cr$_{2}$OAs$_{2}$, although the structure includes an additional perovskite plane, it does not contribute independently to the magnon splitting. The inter-plane couplings are generally weak and frustrated, rendering their direct influence on chiral magnon behavior minimal. Therefore, to reasonably capture the mechanisms underlying the chiral magnon splitting, we restrict our effective spin Hamiltonian analysis to the ILL plane. A full treatment incorporating multi-plane interactions is beyond the scope of this work and will be explored in future studies.
Compared to the other two compounds, Sr$_{2}$CrO$_{2}$Cr$_{2}$OAs$_{2}$ displays the
most pronounced splitting, which exceeds half of the total magnon band width at the $Y/X$ points---an expected outcome given the extreme difference between $J_{2a}$ and $J_{2b}$. 

\section{Conclusions\label{sec:conclusion}}

In this work, we have provided a comprehensive investigation of magnetic phases in ILL systems and practical insights for identifying altermagnetic materials within this structural family. By constructing phase diagrams based on a simple Heisenberg model, we clarified the fundamental mechanisms governing altermagnetism and other complex magnetic phases reported experimentally. To connect theory with experiment, we examined a series of known ILL compounds through density functional theory (DFT) calculations and determined their magnetic ground states. These computational results show good agreement with available experimental data.

A practically important outcome of this study is the identification of a correlation between magnetic ordering tendencies and the $d$-shell electron fillings of transition metal ions, with $d^{3\pm x}$ and $d^{5}$ configurations exhibiting a pronounced propensity toward altermagnetic behavior. 

Finally, using the exchange coupling parameters obtained from DFT calculations, we computed the magnon spectra for selected altermagnetic systems. 
We see that the magnitude of chiral splittings in the magnon dispersion is directly tied to the anisotropy between crystallographically inequivalent $J_{2}$ exchange interactions, as expected.

Of the numerous studied materials we have singled out Sr$_{2}$CrO$_{2}$Cr$_{2}$OAs$_{2}$ as the most exotic and the most promising, and we hope that our work will stimulate further experimental studies.

Overall, our combined theoretical and computational framework offers valuable insights for the prediction and design of new altermagnetic materials and lays the groundwork for future studies exploring the interplay between lattice geometry, electronic configuration, and complex magnetic excitations in frustrated magnetic systems.

\begin{acknowledgments}
The work at George Mason University was supported by the Army Research Office under Cooperative Agreement Number W911NF-22-2-0173, and that at the University of Nebraska by the U.S. Department of Energy (DOE) Established Program to Stimulate Competitive Research (EPSCoR) through Grant No. DE-SC0024284. Most calculations were performed  using resources provided by the Office of Research Computing at George Mason University (URL: https://orc.gmu.edu), funded in part by grants from the National Science Foundation (Award Number 2018631).
\end{acknowledgments}

\bibliography{main}

\end{document}